\documentclass[psfig,graphicx,usenatbib]{aipproc}
\input{aipcheck}
\layoutstyle{6x9}
\usepackage{epsfig,txfonts}
\begin{document}
\title{Knot Based Large Scale Structure Code \\ (complete paper)}
\author{Adrian Sabin Popescu}{address={Astronomical Institute of Romanian Academy, \\ Str. Cutitul de Argint 5, RO-040557
Bucharest, Romania}, email={sabinp@aira.astro.ro}}
\begin{abstract}
In the {\it Dimension Embedded in Unified Symmetry} (D.E.U.S.; \cite{deus}) we made a qualitative description of the way in which we can construct the Large Scale Structure of the Universe from the knot-particle equivalence. Even that we are limited by the lack of computational power implemented on a nonlinear computational architecture needed to conduct this study to its finish, we are still able to give the algorithm to be used in a future simulation, on a, let say, quantum computer.  
\end{abstract}


\maketitle

\section{Code Steps}

\begin{enumerate}
\item Projection of stringy trefoil knot in a plane. The result will be a triangle formed from link crossings (see Fig. \ref{fig1}). The orientation of the knot in space (its angle relative to the projection plane) will be in such a way that the projection link crossings to form a right triangle (see the representation of the FRW Universe bubbles from \cite{deus}, paper II). Because the knot is the instantaneous image of a DEUS structure, the time along its links is constant;

\begin{figure}[htb]
\centering
\leavevmode
\includegraphics[scale=0.7]{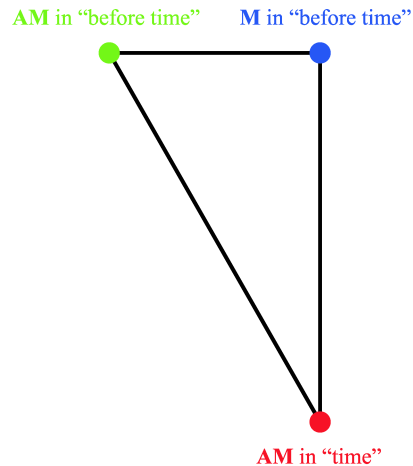}
\caption{Projection of the stringy trefoil knot on a plane. The nodes (link crossings) of this projection must be regarded as the matter (M) and antimatter (AM) in "before time" (before the Big Bang) and "time" Universe presented in \cite{deus}.}
\label{fig1}
\end{figure}

\item Next step is the construction of a network from these triangles on a band, in the projection plane (we reconstruct an unidimensional network of spatial particles), as represented in Fig. \ref{fig2};

\begin{figure}[htb]
\centering
\leavevmode
\includegraphics[scale=0.8]{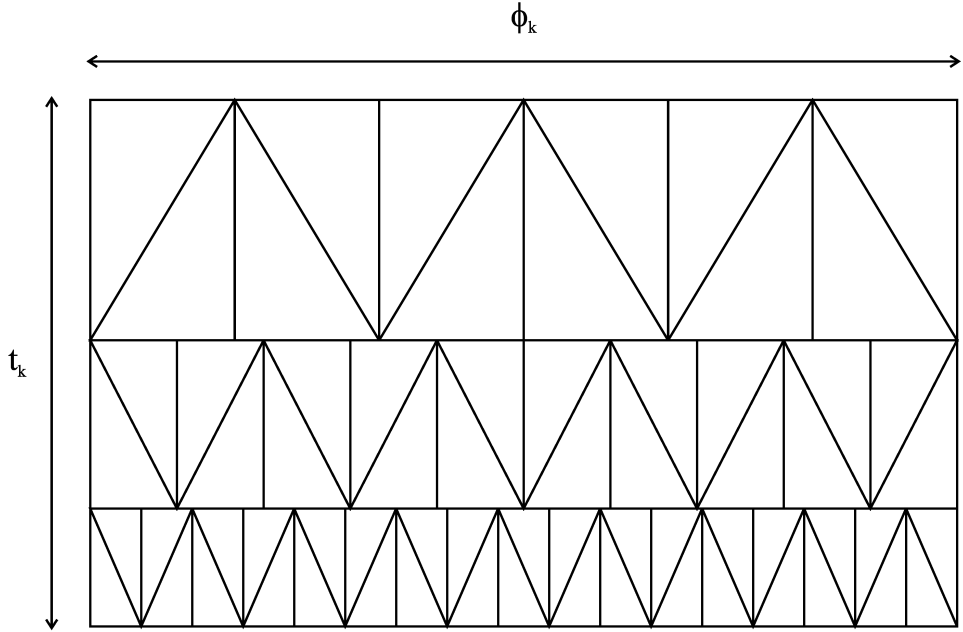}
\caption{Projection of the stringy trefoil knots on a band of coordinates $\phi_{k}$ (helicoid space in DEUS model \cite{deus}) and $t_{k}$ (helicoid time in DEUS model \cite{deus}). After projection and applying the selection rules (see the text) we can convert from ($\phi_{k}, t_{k}$) to ($r, a$) by the \cite{deus} transformations (paper II).}
\label{fig2}
\end{figure}

\item With the above network band we construct banded trefoil knots;
\item We project the banded knots on the walls of a cylinder (we reconstruct a bidimensional network for the space, more or less equivalent with the black hole DEUS self-similarity level);
\item We shape this cylinder as a tube trefoil knot;
\item We construct, on a sphere, a network from the projections of the previous step knots (reconstruct the 3D representation of the space, equivalent with the instantaneous image of the Universe DEUS self-similarity level);
\item By expanding (or contracting) the previous mentioned sphere we will obtain an evolutive representation of the Universe (a space-time representation). The distance between these concentric spheres is given by the height of the tetrahedron from \cite{deus}, paper XIX. The vertex point of imaginary coordinates of the tetrahedron, with the other three vertices (of real coordinates) into the sphere of radius $R_{1}$, will be situated on the sphere of radius $R_{2}>R_{1}$;
\item Further, we construct a cone with the apex on one of the "galaxies" (see, from following selection rules, how we define here the "galaxy") situated on the biggest radius sphere (see Fig. \ref{fig4}). The base radius is $R_{cone}\leq R_{sphere}$. The cone crosses the smaller radius spheres, each one of them containing the final knot network projections (of tube trefoil knots);
\item We project the objects contained in the cones (represented in Fig. \ref{fig4} with red) in their median plane;
\item The final phase is to compare the obtained bidimensional images with the CfA2 redhsift survey \cite{lappa} and SDSS (Sloan Digital Sky Survey; \cite{gott}) results, at the same observational angle. If the simulated data are not fitting the observational ones, we must change the "galaxy" at which we report ourselves and repeat the study starting from step 8.
\end{enumerate}

\begin{figure}[htb]
\centering
\leavevmode
\includegraphics[scale=0.7]{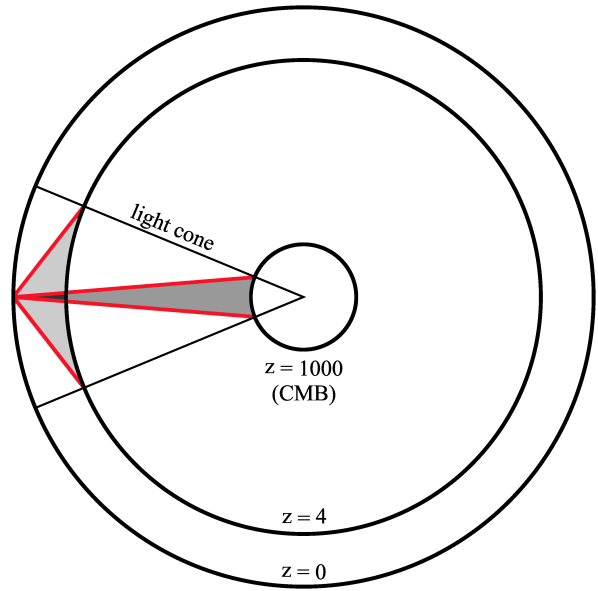}
\caption{Being centered on one of the concentration of points representing a "galaxy", the observer sees the structure of over-imposed knot networks, each contained on a sphere of radius smaller than the observer's one.}
\label{fig4}
\end{figure}

The nodes (link crossings) from the projection of each of the three types of knots (in Fig. \ref{fig1} we have only the nodes of the stringy trefoil knot) represent three matter "hypostases": "time" matter, "before time" antimatter, "before time" matter \cite{deus}. The last two nodes are not visible but influence the obtained structure of the Universe. The "time" antimatter and "time" matter are, in projection, equivalent. This is because if, for example, the "time" matter is considered a node of a right-handed trefoil knot, the "time" antimatter will be the same node but of a left-handed trefoil knot.

\section{Selection rules}

These selection rules will be correctly applied only after the projection on the sphere. All these selection rules will simplify and reduce the time of the simulation. 

\begin{itemize}
\item {\it Object type selection}. \\ We simplify the problem considering (before projection and in five dimensions) that each tube trefoil knot represents a black hole (second level of DEUS self-similarity \cite{deus}), each banded trefoil knot represents an atom or boson (level three of DEUS self-similarity \cite{deus}), and each stringy trefoil knot, the nucleonic or leptonic level (level four of DEUS self-similarity \cite{deus}). 
\item {\it Density selection}. \\ We can take a medium density of atoms and black holes contained in a galaxy. The same density of points will form a "galaxy" in our simulation. Everything that is over this threshold will be a point (a "galaxy") in the simulation.
\item {\it Apex angle (observational angle) selection}. \\ If the structures obtained from simulation present periodicity in the light cone we can reduce the observational angle to one such period.
\item {\it Color code selection}. \\ We will code each node of the trefoil knot with a RGB color (see Fig. \ref{fig1}). Apart of the RGB fundamental color of the nodes, also the domain enclosed between two nodes and a link will be scaled between the two fundamental colors of the nodes (see Fig. \ref{fig3}). Each of this domains will be divided in 100 equal parts. After the projection of one of the knot types (stringy, banded or tube), the {\bf real} matter will be only the points contained into the intersection of such regions (from different triangles in the network) and having 75$\%$ of the summed color of the domain between a "time" and a "before time" node (in physical interpretation, dark energy) and 25$\%$ of the color of the domain between two "before time" nodes (in physical interpretation, dark matter). All the points that do not satisfy this proportion will be excluded from the simulation before the construction of the following knot type (e.g., construction of the banded knot from the band containing the network of stringy knot projections). This selection rule will be applied, in turn, for all the three knot types.

\begin{figure}[htb]
\centering
\leavevmode
\includegraphics[scale=0.8]{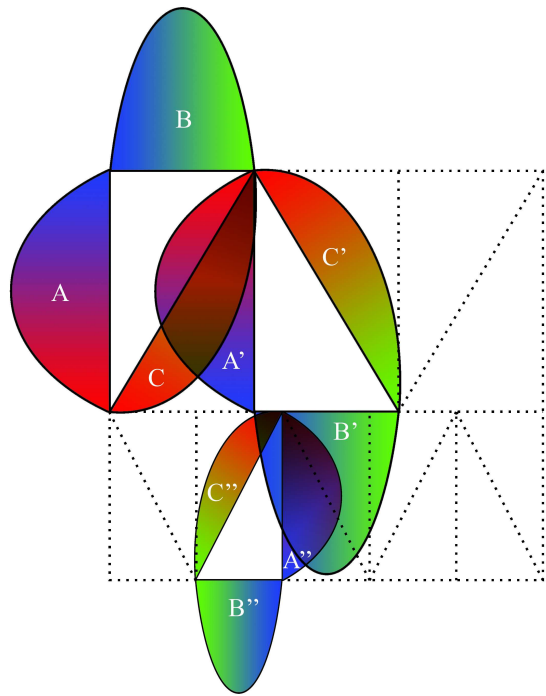}
\caption{Example of the color selection rule applied on a network of stringy trefoil knot projections.}
\label{fig3}
\end{figure}

\end{itemize}

\section{Conclusions}

This paper is a proposal of a new method for the study of the Large Scale of the Universe. At present times, the number of particles needed for this kind of simulation and the highly nonlinear nature of the points distribution (of the implied computations) make impossible applying this method in practice. Even that the given selection rules simplify in big proportion the computational effort and reduce the time of the simulation, it is still not enough. Not only because their computational speed advantage but also, and especially because of this, the possibility of constructing them on a knot based architecture, the best candidate to carry this simulation is the already developing quantum computer. It is appealing that, instead of working on a nonlinear distribution with a linear "thinking", normal, computer, to use in the process a computer constructed and "thinking" in the same way as the problem to be solved (the Universe as a matter of speaking).

\section{Acknowledgements}
I want to express all my gratitude to Stefan Sorescu whose computer drawing experience was truly valuable at the creation of this paper figures.

\end{document}